# Design Principle of Gene Expression Used by Human Stem Cells; Implication for Pluripotency

Michal Golan-Mashiach[1,2*], Jean-Eudes Dazard[1,3*], Sharon Gerecht-Nir[4], Ninette Amariglio[5], Tamar Fisher[5], Jasmine Jacob-Hirsch[5], Bella Bielorai[5], Sivan Osenberg[4], Omer Barad[2], Gad Getz[2], Amos Toren[5], Gideon Rechavi[5], Joseph Eldor-Itskovitz[4], Eytan Domany[2], David Givol[1§].

[*]These authors contributed equally to this work.

[1]Department of Molecular Cell Biology, [2]Department of Physics of Complex Systems, Weizmann Institute of Science, Rehovot, 76100, Israel, [3]Department of Epidemiology and Biostatistics, School of Medicine, Case Western Reserve University, Cleveland, OH 44106, USA, [4]Department of Obstetrics and Gynecology and Biotechnology Interdisciplinary Unit, Rambam Medical Center, Faculty of Medicine, The Technion, Haifa, Israel; [5]Department of Pediatric Hematology-Oncology, Chaim Sheba Medical Center and Sackler School of Medicine, Tel-Aviv University, Tel-Aviv 52621, Israel.

[§]To whom correspondence should be addressed. E-mail: david.givol@weizmann.ac.il tel. 972-8-9343634 fax. 972-8-9344125

**Running Title**: Design Principle Used by Human Stem Cells

**Abstract**

Human embryonic stem cells (ESC) are undifferentiated and are endowed with the capacities of self renewal and pluripotential differentiation. Adult stem cells renew their own tissue, but whether they can trans-differentiate to other tissues is still controversial. To understand the genetic program that underlies the pluripotency of stem cells, we compared the transcription profile of ESC with that of progenitor/stem cells of human hematopoietic and keratinocytic origins, along with their mature cells to be viewed as snapshots along tissue differentiation. ESC gene profile show higher complexity with significantly more highly expressed genes than adult cells. We hypothesize that ESC use a strategy of expressing genes that represent various differentiation pathways and selection of only a few for continuous expression upon differentiation to a particular target. Such a strategy may be necessary for the pluripotency of ESC. The progenitors of either hematopoietic or keratinocytic cells also follow the same design principle. Using advanced clustering, we show that many of the ESC expressed genes are turned off in the progenitors/stem cells followed by a further downregulation in adult tissues. Concomitantly, genes specific to the target tissue are upregulated towards matured cells of skin or blood.

**Introduction**

Human embryonic stem cells (ESC) exhibit self-renewal and pluripotential differentiation (1). Adult tissue specific stem cells renew their own tissue and may also exhibit plasticity by trans-differentiation into cells of other tissues. Research on stem cells attracted recently the attention of the public as well as the scientific and medical community because of their potential for cell replacement therapy with the option to personalized stem cells produced by nuclear transfer (2). The gene expression profile of ESC can be highly informative with regard to their self renewal and pluripotency properties. Although some genes that are important for maintaining the undifferentiated state of ESC were identified (e.g. *NANOG* (3) *OCT4* (4)) the genetic program and design principle that underlies the pluripotency of stem cells is not well understood.

Recent work used microarray analysis to characterize the gene expression profile of several human ESC cell lines and to compare it to that of closely related tumors like embryonal carcinoma and seminoma or to a variety of somatic cell lines (5). The results demonstrated the high similarity in expression profiles between the various ESC lines, the close relationship of ESC profiles to those of embryonal carcinoma or seminoma and a distant relationship to somatic cell lines. They identified 895 genes that were expressed in human ESC at higher level than in control samples. Another study compared ESC profiles to a mixture of RNA obtained from various adult tissues (universal RNA) assuming that it represents a reference point for differentiated adult cells (6). In this study a group of 92 genes were found to be enriched in expression in all six human ESC tested and were considered a molecular signature of the stemness properties of ESC.

We wished to compare the expression profile of human ESC to that of defined adult tissues and their progenitors, which are related to the much sought for minor population of adult tissue stem cells. For this study we chose blood and epidermis because they exhibit tissue renewal throughout life, and the differentiated cells of such tissues have a short life span and do not divide. We isolated hematopoietic cells' progenitors from whole peripheral or cord blood by adsorption to anti CD133 (7), and the keratinocytes' progenitors by taking advantage of the differential expression in the hemidesmosomal integrins β1 and β4 of epidermal keratinocytes (8, 9). The expression profiles of ESC and of such cell populations represent snapshots of cell states, from embryonal stem cells through tissue progenitor to differentiated cells; comparison of such a sequence of expression profile may help to elucidate the gene expression pattern which is unique for embryonic stem cell and may be responsible for their properties.

We found that ESC express significantly more genes than adult tissue cells; many of these genes are downregulated upon differentiation to adult tissue cells. The down-regulated genes that are unique in their expression to ESC include genes that were identified previously as responsible for pluripotency and self-renewal, such as *NANOG* (10) *OCT4* (4) and others. Surprisingly many of the down-regulated genes are known to be expressed in a variety of other adult tissues. On the other hand, some other genes that are specific to the target tissue (e.g. epidermis, blood) are up-regulated with respect to their expression in ESC. We suggest the hypothesis that in order to maintain their pluripotency, many differentiation options are primed in ESC by promiscuous gene expression **just in case** some of them will be needed for the target tissue towards which the ESC is instructed to differentiate.

This represents a design principle of stem cells that accounts for both self renewal and pluripotency. A prime candidate for pluripotential differentiation is the parsimonious "just in time" strategy; expressing genes only when needed, i.e. at the moment of commitment to a particular differentiation path. The opposite extreme is the seemingly more wasteful "just in case" strategy, which keeps a wide repertoire of expressed genes, to be present in case a particular path is selected. Our results indicate that stem cells keep thousands of non specific genes expressed ("just in case") that are selected for upregulation or quenched (if not needed) upon differentiation.

**Methods**

**Isolation and characterization of cell samples used for microarrays analysis**

Human ESC were obtained from H9.2 cloned line, a derivative of H9 that was previously isolated from the blastocytes' inner cell mass (11, 12). Several surface markers typical of primate ESC (11, 12) were examined using immunofluorescent staining. The H9.2 cells were found to be positive for the surface markers *SSEA4*, *TRA-1-60* and *TRA-1-81*. RT-PCR analyses further showed the expression of the pluripotency markers *NANOG* and *OCT4*.

Similar to other ESC, once removed from its feeder layer and cultured in suspension (13), the H9.2 line formed Embryoid Bodies (EB), including cystic EB. Histology sections revealed different cell types within these EB. The injection of undifferentiated H9.2 cells into the hindlimb of SCID-beige mouse resulted in the generation of a teratoma which possessed representatives of the three germ layers.

Human keratinocytes were obtained from 12 pooled neonatal foreskins as previously described (14) or from primary culture of normal human keratinocytes. To obtain progenitor/stem cells from keratinocytes, we took advantage of their differential expression of the integrins β1 and β4 (8, 9). The keratinocytes were separated into 3 fractions by differential adsorption on native collagen IV, a fundamental epidermal basement membrane component and a ligand of integrins β1 and β4. Rapidly adherent cells on collagen-IV coated plates were obtained after 0.5h of incubation and harvested as keratinocytes stem/progenitor cells (KSPC). Unadsorbed cells were re-plated on new collagen-IV coated plates and the cells adsorbed after 24h were harvested and collected as transit amplifying cells (TAC) whereas the unadsorbed cells after 24h were collected as differentiated keratinocytes (KDC). Cells from the three fractions were analyzed for clonogenicity and expression of known markers (14). Fig. 1 demonstrates that KSPC show the highest clonogenicity potential and the expression of markers known to be characteristic to epidermal stem cells, like *P63* and β–catenin as well as the absence of keratinocytes' differentiation markers.

Human progenitors for hematopoietic cells were obtained from pools of cord blood collected after placental separation and from peripheral blood collected by pheresis from adult normal donors primed with G-SCF for stem cells mobilization. Enriched hematopoietic stem/progenitor cells were isolated from those preparations, using the anti CD133 magnetic beads separation system (Miltenyi). The non-selected cells were termed hematopoietic differentiated cells (HDC). The yield of selected CD133 positive cells was 0.14% for cord blood and 0.7% for peripheral blood and the isolated cell populations were 80-85% positive for CD133 as assayed by FACS. Evidently the CD133+ cells contain CD34+ cells and a significant amount (~25%) of CD34+, CD38+ cells. Although hematopoietic stem cells are characterized by CD38- markers it is known that even such a population (CD34+, CD38-) is heterogeneous. The human CD133+ cells are considered a population of hematopoietic stem/ progenitor cells (15) and we defined them as HSPC. To support this contention we compared the results of enriched expressed genes of HSPC in our samples with that of human cord blood samples characterized by LIN- CD34+ CD38- reported by Ivanova et al (16). Out of 2500 Probe Sets (PS) with the highest fold change reported by them (16)1198 PS ((48%), 1055 genes) are common with our list derived from the CD133+ cells. The common genes include *CD34*, *AC133*, *BMI1*, *TIE*, *HOXA10*, *FZD6*, *SOX4*, *TCF3*, *RUNX1*, *FLT3*, *LMO2*, *SCA1*, *MEIS1, MPL, TEK, SCA1, MDR1* and others. Theses genes are known to be important for the maintenance of hematopoietic stem cells. Hence, our CD133 isolated hematopoietic cells (HSPC) represent a similar population to that of human LIN- CD34+ CD38- hematopoietic stem cells reported by Ivanova et al (16) .

**RNA extractions and microarray.** Gene expression was measured by hybridization to Affymetrix U133A GeneChip® DNA microarrays, containing ~22,000 probe sets (PS). RNA extractions were performed by the RNA isolation reagent TRIzol®, and RNA integrity was assessed by gel electrophoresis. 10 μg total RNA was used to prepare biotinylated target cRNA according to Affymetrix™ procedures.

**Bioinformatics.** For probe-level data analysis, we tried several methods of probe set summarization: MBEI (http://www.biostat.harvard.edu/complab/dchip/), RMA (http://www.bioconductor.org) and MAS 5.0. The reported results did not depend on the particular method used and MAS 5.0 expression values and detection calls were kept. Expression levels < 30 were thresholded to 30 and $\log_2$ was taken to generate the final gene expression matrix. For clustering analysis, a new very efficient version (O. Barad, M.Sc. thesis, Weizmann Inst. of Science, unpublished) of the Super Paramagnetic Clustering method (17) was used to cluster the data. Prior to clustering, the expression levels for each PS were centered and normalized.

**Tissue Classification of Expressed genes.** The MAS 5.0 signal intensities of the GNF dataset (http://expression.gnf.org/cgi-bin/index.cgi) of Su et al. (18), that contain expression data for 21 tissues were converted into quantile scale (19) as follows: the expression levels for each gene in all tissues were divided into 10 equal bins for signal levels above 50. Signal below or equal to 50 was termed zero quantile. The expression level of each gene in all tissues analyzed is represented by the quantile numbers between 0 and 10. The maximum difference between two neighbouring values in the sorted quantile vector is defined as the "gap" index. The "gap" criterion was used to convert the expression profiles into a binary form. For each probe set, we first checked whether there is a gap with value of 3 or higher. Tissues in which expression was detected above the threshold gap were interpreted as over-expressing the gene and marked as 1. The rest of the tissues were referred to as under-expressing the gene and marked as 0. In this manner each expression profile of a PS over the tissues was classified to a particular binary pattern. The probe sets were divided into 3 groups: (i) probe sets with quantile numbers = 0 for all tissues represent genes absent in all tissues (ii). Probe sets with only one or two tissues above the gap represent tissue-specific or "tissue-affiliated" genes, and (iii) probe sets with over-expression in more than 3 tissues are called genes "expressed in multiple tissues". For example, if a gene is expressed in the various tissues with all the levels, ranging between 1 and 10, the gap is 1 and the gene is labelled as "expressed in multiple tissues". If the gene is expressed at a level of 10 in one tissue and a level of 2 in all other tissues, the gap is 8 and the gene is tissue specific for the tissue with the expression level of 10.

**Results**

**ES Cells Express More Genes than Differentiated Cells.** To assess the changes of gene expression during differentiation, we studied 17 samples: 3 ESC, 4 HSPC and 4 HDC, 3 KSPC and 3 KDC. We wished to know whether ESC and tissue adult cells express a similar or different number of genes. The gene expression results of the ESC and adult differentiated cells (10 samples) were analyzed as follows: out of 22,215 probe sets (PS) on the chip, 15,918 PS with at least one "present" call, obtained from MAS 5.0, were kept. The $\log_2$-transformed expression levels for each PS were centered and normalized. The values of each gene were between -1 and +1, where 0 is the average expression of this gene in these samples. A value of 0.3 in a particular sample means a significant expression above the average in this sample. This criterion was used by others (20) and we also defined a PS with a centered and normalized value above 0.3 as highly expressed. Table 1 shows that 4550 PS satisfy this criterion in ESC; this number is significantly higher than the number of similarly identified genes that are highly expressed genes in the adult differentiated state (~3000 PS).

We wanted to classify these genes in order to know whether they are specific to the embryonic state and for this we studied the expression levels of these 4550 PS in a wide group of adult tissues. This was done using the GNF dataset (http://expression.gnf.org/cgi-bin/index.cgi) described by Su et al (18) who studied the expression level of 21 human tissues (see Methods). The 4550 PS were converted from U133A PS to U95A PS and then were classified into 3 groups as follows: (a) Genes that are highly expressed in ESC but not in any adult tissue that was analyzed (18) (low expression is considered less than a signal of 50 as calculated by the affymetrix MAS 5.0 software) (b) tissue-specific genes, expressed in 1-2 tissues (c) genes expressed in multiple adult tissues (> 2 tissues). This analysis was performed by enumeration of graded expression levels for each PS in every tissue (see Methods).

We found that from 4550 PS that are expressed in ESC, 700 PS are not expressed in any of the adult tissues studied by Su et al (18). This number is similar to that obtained by others (5). Most (3300) of the remaining PS are expressed in multiple adult tissues and approximately 1000 PS are tissue specific genes (fig 2). Examples for tissue-specific genes expressed in a variety of tissues can be seen in table 2.

**Marked Down-Regulation of Expressed Genes Upon Stem Cell Differentiation**. The above results suggest that upon differentiation from ESC to adult cells there is a marked down regulation of expressed genes. To test this we analyzed two groups of samples which may represent virtual steps in the differentiation pathways from ESC to adult cells:

(A) Hematopoietic (H) pathway; ESC → HSPC → HDC (3+4+4 samples), and

(B) Keratinocytic (K) pathway; ESC → KSPC → KDC (3+3+3 samples).

It is clear that the three cell stages depicted in the above pathways do not represent real differentiation but comparison of snapshots of cell stages derived from different sources. Nevertheless they can be considered as representatives of different stages between embryonic stem cells and mature tissue cells. For each group the genes were filtered using *ANOVA* (21) and false discovery rate (FDR) (22) was controlled at 0.05. This left 8290 PS (6293 genes) that vary significantly over the three kinds of cell states in group (A) and 5432 (4301 genes) for group (B); There are two sources to this difference. The first is the different numbers of samples in the two groups. Discarding two samples from group (A) and repeating the ANOVA with 3+3+3 samples yields about 7000 PS (instead of 8290) that pass with FDR of 0.05. The further reduction to 5432 PS is due to the fact that the within-group variance of the genes in the K pathway significantly exceeded that of the same groups in the H pathway. We present in Figs. 3A and 3B the expression levels of the significantly varying PS. The black line shows the expression level of 8290 PS in ESC from highly expressed PS (left side) to the lower expressed PS (right side). The green dots show the expression of the same genes in HSPC (Fig. 3A) and KSPC (Fig. 3B) and the red dots represent the expression of the same genes in adult cells (HDC Fig 3A, KDC Fig 3B). The data showed that ESC (black line, Fig. 3) express many genes at a higher level than the other cells analyzed and the majority of transcripts exhibited marked downregulation along the differentiation pathway: 4392 PS (3483 genes) were downregulated as cells differentiate from ESC to HDC, this was accompanied by up-regulation of a smaller group of 2638 PS (1988 genes), with low expression in ESC and high in the HDC. A similar pattern was seen in the keratinocytic pathway (Fig. 3B).

**Clustering analysis shows distinct stem cells specific genes.** We clustered (17) the samples of groups A (8290 PS) and B (5432 PS) separately, to identify distinct differentiation-related variations of the expression profiles, and to assign genes to clusters of similar patterns of expression. Fig. 4 depicts the expression matrix after clustering of the genes in the H (Fig. 4A) and K (Fig. 4B) pathways. Six clusters are clearly shown: clusters H1, H2 and H3 contain ESC genes that were down-regulated along differentiation in the H pathway and K1, K2, K3 were downregulated in the K pathway. Clusters 4 and 5 contain genes that were upregulated along the virtual differentiation pathways (H or K). Clusters 6 contain genes expressed only in adult stem/progenitor cells, whereas clusters 3 contain genes that show expression only in ESC. Clearly, ESC and adult stem progenitor cells have different gene expression profiles. The lists of genes in each of the clusters are given in tables S2-S13 in supplementary data (http://www.weizmann.ac.il/physics/complex/compphys/downloads/michalm/).

Table 3 presents selected genes that were previously shown to be typical of one of the cell stages, and belong to one of the six clusters in Fig. 4. Clusters 1 and 2 contain genes that are common to ESC and adult stem/progenitors cells and therefore may represent the "stemness" genes as previously defined (16, 23). It should be noted, however, that many genes, well known to be markers for undifferentiated ESC or related to ESC self-renewal [e.g. *NANOG, POU5F1* (*OCT4)*, *SOX2*, *FOXH1, TDGF1 (Cripto), LeftyA & B*, *Thy1* (3, 4, 24-26) – Table 3] belong to clusters H3 and K3 (Fig. 4), and thus are suppressed in HSPC or KSPC and seem to be unique to the ESC. Their roles are apparently taken over in the tissue stem/progenitor cells by those of clusters H6 or K6, which show expression only in the stem/progenitor cells, and indeed contain genes known to be essential for the self-renewal and tissue differentiation and development [e.g. *TP73L (p63)*, *ITGB4* and *BNC* for skin (27-29), and e.g. *BMI1*, *CD34*, *TIE*, *KIT*, *TAL1* (*SCL*), and *RUNX1* for blood (30-32) – Table 3]. It is of interest that the genes present in cluster 6 (H6, K6, Fig. 4) show exclusive expression in progenitor cells of either blood or epidermis. For example *CD34* and *KIT*, known markers of hematopoietic stem cells are included in this cluster. The recently identified *BMI1* that function to maintain hematopoietic stem cells pluripotency and *RUNX1* a transcription factor that regulates haematopoiesis and involved in leukaemia are also included in H6. In the KSPC (epidermis keratinocyte pathway) a notable example is p63 that was shown by Knock-out to be essential for skin development (33, 34). Hemidesmosomal and adherence junction components such as integrin β4 and *BPAG1* and *BPAG2* keep keratinocyte attached to the epidermal basement membrane, thereby participating in the maintenance of the undifferentiated state of keratinocyte stem cells. The fact that *P63*, integrin β4, *BPAG1* and *BPAG2* also belong to cluster K6 indicates that this cluster contains genes related to progenitor keratinocyte cells. Basonuclin a zinc finger protein associated with increasing transcription of rRNA and epithelial proliferation (35) is also included in K6. We can conclude that the cell population isolated by us as representative of stem/progenitor cells indeed exhibit many characteristic genes for the young immature epidermis progenitors and they are not expressed in the adult stage of this tissue. On the other hand the clusters H4, H5 and K4, K5 contain genes that are markedly upregulated along the differentiation pathways towards the target tissue (blood or epidermis). These groups of genes include blood cells specific genes like *IGG, MHC, Haptoglobin* and *Fc* to mention only a few. In the keratinocytes pathway completely different set of genes is upregulated in K4 and K5 and include epidermal differentiation markers like Loricrin, Filaggrin and keratins 10 and others (Table 3).

**Search for a core of ESC expressed genes, some of which may represent an ESC signature.** The unique gene list of clusters 3 (H3 and K3, Fig. 4 and table 3) may contain signature genes for ESC since they are expressed only in ESC when clustering is performed on either of the pathways. To search for a common core of ESC genes we intersected the H3 and K3 gene lists. This resulted in a list of 179 genes that contain genes highly enriched in ESC relative to differentiated and stem/progenitors of both blood and epidermis, but not necessarily relative to other tissues. These genes may be regarded as candidates for serving as embryonic stem cell signature. However, checking the behaviour of these genes in the data of Su et al (18), we discovered that 98 genes (out of the 179) are expressed in one or more adult differentiated tissue. We also checked this list of genes against available data on expression of their orthologs in murine adult tissue stem cells. Of the remaining 81 genes 15 are expressed either in other adult stem cells (36, 37) or in other adult tissue as found by checking them individually in the Genecards dataset (http://bioinfo.weizmann.ac.il/cards/index.shtml), leaving 66 candidate genes (table S1 in supplementary data http://www.weizmann.ac.il/physics/complex/compphys/downloads/michalm/). This list may shrink further if more adult stem cell lineages are studied. This list includes most of the well known genes that were analyzed in various ESC systems and shown to be essential for pluripotency and self-renewal, like *OCT4, LIN28, TDGF1, LeftB, SOX2,* and others. Twenty significant genes from this list are presented in table 4. Among these genes is *OCT4*, a transcription factor that is expressed in ESC and down regulated upon differentiation (38). Similarly, *NANOG* is another transcription factor that is involved in pluripotency and suppression of differentiation and *NANOG* deficient ESC show limited pluripotency and produce only endoderm like cells (3 , 10). *TDGF1* or *cripto* is an autocrine growth factor and stimulates cell proliferation at the expense of differentiation. *TDGF1*, when deficient, is embryonic lethal at post gastrulaion due to failure of morphogenesis (39). *LEFTB* is a *TGFβ* related protein that regulates left-right early embryonic patterning (40). The transcription factor *SOX2* that regulates *FGF4* expression, was found in all human ESC and is also a marker of neural progenitors. Frizzled (*FZD*), is a receptor for *WNT* and part of the *WNT*/β-catenin pathway. β-catenin activates the *TCF* transcription factor leading to proliferation and inhibition of differentiation (41, 42). Constitutive expression of *SOX2* maintains the progenitor state and inhibits neuronal differentiation (43). It may operate together with *OCT4* to specify the three body layers at implantation (24). The list includes also *CDC6* that is required for initiation of DNA-replication (44). Noteworthy in this list are genes that encode for protein involved in remodelling of chromatin. For example, *SMARCA1* an homolog of the yeast general transcription activator of the *SWI/SNF* family members which known to be involved in chromatin remodelling (45). This complex includes also members of *SWI2/SNF2* family and histone deacetylase (*HDAC*). This suggests that a significant part of the regulation of gene expression in ESC is by epigenetic changes that regulate chromatin alteration.

**Discussion**

In this study we analyzed the expression profile of human ESC line 9.2 and compared it to the expression profiles of adult tissues - differentiated cells (blood and keratinocytes) as well as their progenitor cells. This comparison, based on clustering analysis, allowed the construction of virtual differentiation pathways from ESC to the differentiated cell of the particular tissue and to identify the genes whose expression may constitute a profile signature for the various steps in this pathway. The results indicate that cluster 3 (table 3) contains genes well-known for their role in the properties of ESC that are absent in adult stem cells. Cluster 6 (table 3) includes genes that are essential for the function of blood and skin stem cells but absent from ESC. This may cast doubt on the notion that all stem cells share a common genetic signature (16, 23).

An important question concerning the gene expression profile of ESC is to try to define the gene list which is ESC-specific or highly enriched and may mark the signature of ESC. These genes should be characterized by their contribution to self-renewal, pluripotency and control of differentiation. A recent study (6) described a list of 92 genes that are common to six hESC lines and not expressed (at least 3 fold lower) in a mixture of RNA (universal RNA) which represents many adult differentiated tissues. Five out the 92 genes were absent from our Affymetrix chip. Of the remaining 87, we found 66 (76%) in clusters H1, H2 and H3 (Fig. 4), i.e. in clusters of genes with high expression in ESC and down-regulation in HDC. Only 26 of these 66 genes belong to cluster H3 and 40 to cluster H1+H2. This does not contradict (6); note that their 92 genes were selected by comparing expression in human ESC with a universal RNA that was derived from adult differentiated tissues, but not from adult tissue stem cells. Their list of 92 includes relatively many genes that are expressed in tissue stem cells (that belong to clusters 1 or 2). Only 9 of the 26 genes (marked by X in table 4) genes appear also in our list of 66 candidate ESC signature genes. In another study, significantly enriched genes in human ESC were identified, including specific genes coding for receptors (5) . We found that all of the most significant 25 receptor genes reported in that work and 15 out of the 25 most significant enriched genes (5) were present in clusters 1, 2 and 3 (Fig 4). Among the most significant receptors we found the *WNT* receptor Frizzled 7 (*FZD7*) and *BMP* receptor that has been recently shown to induce human ESC to differentiate into trophoblast (46).

An interesting result of our data was the decrease of complexity of gene expression upon differentiation from ESC to adult cells (table 1). The clustering results also showed that overall in the hematopoietic pathway 4392 PS (3483 genes) were downregulated and 2638 PS (1998 genes) were upregulated, while in the keratinocyte pathway 3417 PS (2758 genes) were downregulated and 1423 PS (1115 genes) were upregulated. The massive downregulation is consistent with the "*just in case*" design principle underlying pluripotential differentiation. We suggest the hypothesis that in order to maintain their potential for pluripotency, ESC "keep their options open" by promiscuous gene expression, maintaining thousands of genes at intermediate levels, and selecting only a few for continuous expression that are needed for differentiation to the target tissue. The rest will be down-regulated upon commitment to a cell fate for which they are not needed. This is a selection model of gene expression during differentiation and the down-regulation of genes that are not needed is required for establishing the differentiated state. It is very likely that this strategy is made possible by maintaining an open chromatin structure at the stem cell stage and epigenetic modification upon differentiation (47, 48). A similar model was previously proposed in the case of hematopoietic stem cell differentiation on the basis of expression of erythroid or granulocyte markers in the progenitor cell prior to commitment (49) and recent work extended this also to the analysis of genes expression by microarrays in the hematopoietic system (20, 50). Our study demonstrates the generality of this model and extends it to human ESC and adult stem cells at the level of global gene expression.

**Acknowledgements**

This work was partly supported by the Arison Family Foundation, the Yad Abraham Research Center for Cancer Diagnosis and Therapy, the Clore foundation, the German-Israeli Science Foundation (GIF), and grants from the Israeli Academy of Science, Minerva and the Ridgefield Foundation. We thank John R. Walker (Novartis, San Diego) for providing us with the GNF data set. G. R. holds the Shapiro chair in hematologic malignancies in Tel-Aviv University. E.D. is the incumbent of the H.J. Leir Professorial Chair.

**Figure legends**

**Figure 1 Clonogenicity assay and expression of epidermal specific markers of keratinocyte fractions.** (**A**) Clonogenicity assay: after selection of isolated keratinocytes from human epidermis on type IV collagen (see Methods), 2000 cells of each fraction were plated per well (in triplicate), and after two weeks in culture, keratinocyte colonies were scored. Numbers represent averaged cFu / abortive colonies (8, 14). (**B**) Western blot analysis of specific markers (27) in keratinocyte fractions isolated as in A. KSPC, Keratinocyte Stem/Progenitor Cells; TAC, Transit Amplifying Cells; and KDC, Keratinocytes Differentiated Cells.

**Figure 2. Classification of tissues expression for the ESC expressed genes.**
The list of 4550 PS (table 1) that are highly expressed in ESC were classified according to their expression in 21 tissues based on the GNF dataset (see methods) into 3 groups : (**A**) ESC specific genes that are not expressed in any adult tissue (expression signal less than 50 in all tissues.) (**B**) Tissue-specific genes expressed in 1-2 tissues C, Genes expressed in multiple adult tissues (> 2 tissues).

**Figure 3. Expression levels of probe-sets (PS) that vary significantly between ESC, adult stem cells and differentiated cells.** The PS were ordered according to their ESC expression levels, marked by black circles that form a line. The expression levels in HSPC or KSPC are indicated by green dots and in HDC and KDC by red dots. (**A**) Expression levels of 8290 PS that vary between ESC, HSPC and HDC. (**B**) Expression of 5432 PS that vary between ESC, KSPC and KDC. Only PS with P-values that passed ANOVA at an FDR level of 0.05 were plotted (21). In (A) 4392 PS (3483 genes) are down-regulated and 2638 PS (1988 genes) are up-regulated with

differentiation. In (B) 3417 PS (2758 genes) and 1423 PS (1115 genes) are up- and down-regulated, respectively.

**Figure 4. Clustering analysis of PS expression levels in hematopoietic and keratinocytic pathways.** The expression levels of the PS taken from Fig. 3 were centered and normalized and the PS were reordered according to the dendrogram produced by the SPC algorithm (17). (**A**) Expression matrix of 8290 PS in ESC, HSPC, and HDC. (**B**) Expression matrix of 5432 PS in ESC, KSPC and KDC. (**C**) Overlaps between the related 6 clusters were calculated relatively to keratinocyte clusters.

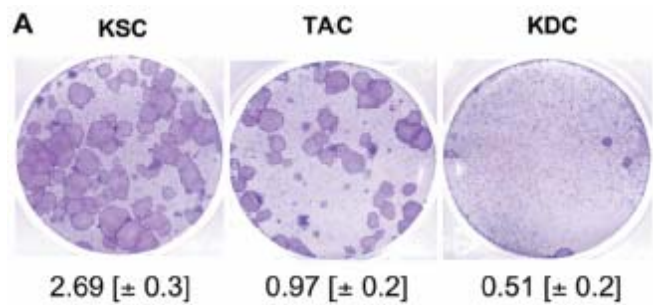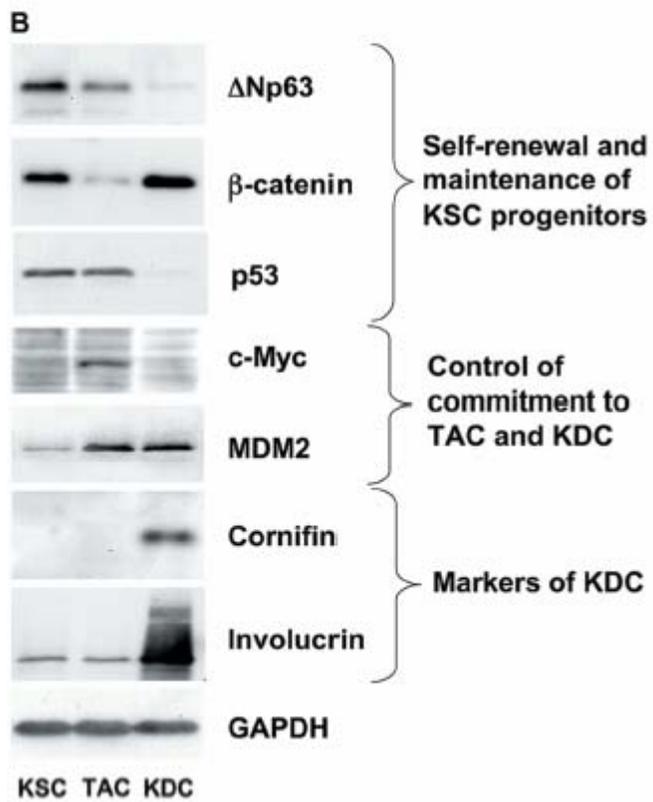

Figure 2

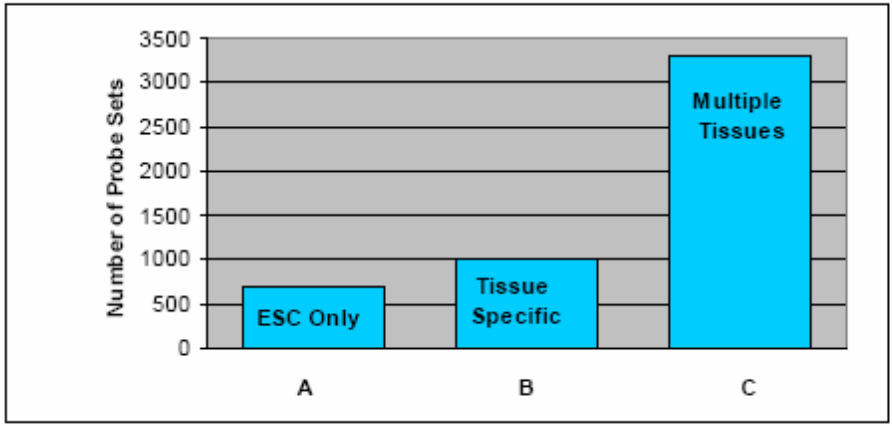

Givol_Fig3

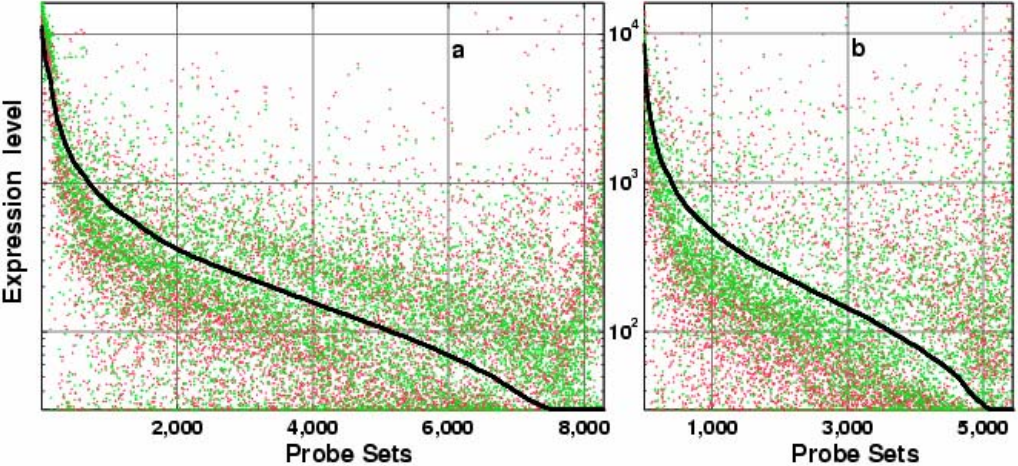

Givol_Fig4

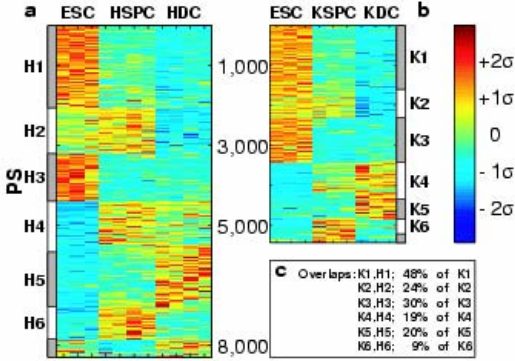

| Table 1. Distribution of Gene Expression in Embryonic Stem Cell and Adult Cell[i] | | | |
|---|---|---|---|
|  | ESC | KDC | HDC |
| Probe sets on the chip | 22,215 | 22,215 | 22,215 |
| Probe sets with at least one "Present" call | 15,918 | 15,918 | 15,918 |
| Probe sets with normalized value more than 0.3 | **4550** | **2982** | **3217** |

---

[i] The numbers in the third row were obtained as follows: each probe set was centered and normalized over the 10 samples (ESC -3, KDC-3, HDC-4) used for the analysis. Probe sets with a value above 0.3 in the samples of each group (ESC, KDC and HDC) were selected and presented.

| Table 2. Selected Tissue Specific Genes Expressed in Human Embryonic Stem Cells as obtained from Fig 2 by checking the GNF dataset. | | | |
|---|---|---|---|
| Symbol | Title | Tissue | ESC Signal |
| CD24 | CD24 antigen | Blood | 4818.20 |
| GRN | Granulin | Blood | 728.43 |
| MMD | monocyte to macrophage differentiation-associated | Blood | 956.80 |
| COL1A1 | collagen, type I, alpha 1 | Bone | 2167.10 |
| CHN1 | chimerin (chimaerin) 1 | Brain | 357.90 |
| CNTNAP2 | contactin associated protein-like 2 | Brain | 315.60 |
| DBN1 | drebrin 1 | Brain | 744.40 |
| DCTN1 | dynactin 1 | Brain | 311.60 |
| KRT10 | keratin 10 | Skin | 883.10 |

| | | | | |
|---|---|---|---|---|
| OPHN1 | oligophrenin 1 | | Brain | 457.53 |
| RTN3 | reticulon 3 | | Brain | 670.03 |
| TMSNB | thymosin, beta, identified in neuroblastoma cells | | Brain | 1550.60 |
| YWHAQ | tyrosine 3-monooxygenase/tryptophan | | Brain | 3343.10 |
| ZIC1 | Zic family member 1 (odd-paired homolog, Drosophila) | | Brain | 1011.53 |
| DDR1 | discoidin domain receptor family, member 1 | | Brain + Lung | 694.03 |
| RTN4 | reticulon 4 | | Brain + Testis | 1859.57 |
| COL11A1 | collagen, type XI, alpha 1 | | Cartilage | 158.97 |
| PIM1 | pim-1 oncogene | | Hematopoietic | 255.17 |
| CRYZ | crystallin, zeta (quinone reductase) | | Lens | 186.73 |
| APOA2 | apolipoprotein A-II | | Liver | 387.20 |
| APOC1 | apolipoprotein C-I | | Liver | 387.47 |
| NGFRAP1 | nerve growth factor receptor | | Ovary + Testis | 5376.73 |
| FHL2 | four and a half LIM domains 2 | | Skeletal Muscle | 221.47 |
| FLOT2 | flotillin 2 | | Skin | 632.37 |
| CALD1 | caldesmon 1 | | Smooth Muscles | 629.40 |
| MYL9 | myosin, light polypeptide 9, regulatory | | Smooth Muscles | 669.37 |

| Table 3. Selected genes identified in clusters of Fig. 4 that are known to be important in the various cell stages | | | | | | | |
|---|---|---|---|---|---|---|---|
| | Hematopoietic clusters | | | | Keratinocytic clusters | | |
| Clust. | Identifier | Symbol | Short Name | Clust. | Identifier | Symbol | Short Name |
| H1 | X52078.1 | TCF3 | transcription factor 3 | K1 | BG393795 | TCF3 | transcription factor 3 |
| | NM_014366.1 | NS | nucleostemin | | AK026674.1 | TCF4 | transcription factor 4 |
| | BF510715 | FGF4 | FGF4 | | BF510715 | FGF4 | FGF4 |
| | L37882.1 | FZD2 | frizzled 2 | | U91903.1 | FRZB | frizzled-relat. prot. |
| | NM_000435.1 | NOTCH3 | Notch 3 | | NM_001845.1 | COL4A1 | collagen, type IV, $\alpha 1$ |
| | AF029778 | JAG2 | jagged 2 | | AK026737.1 | FN1 | fibronectin 1 |

|    | Accession    | Symbol  | Name                  |    | Accession    | Symbol  | Name                   |
|----|--------------|---------|-----------------------|----|--------------|---------|------------------------|
|    | AL556409     | GAL     | galanin               |    | NM_021953.1  | FOXM1   | forkhead box M1        |
|    | NM_005842.1  | SPRY2   | sprouty 2             |    | AL556409     | GAL     | galanin                |
|    | NM_001903.1  | CTNNA1  | α catenin 1           |    | NM_005359.1  | MADH4   | SMAD4                  |
|    |              |         |                       |    |              |         |                        |
| H2 | AK026674.1   | TCF4    | transcription factor 4| K2 | BC004912.1   | BPAG1   | bullous pemph. ag.1    |
|    | NM_001331.1  | CTNND1  | δ catenin 1           |    | NM_003798.1  | CTNNAL1 | α catenin like 1       |
|    | M87771.1     | FGFR2   | KGF receptor          |    | NM_022969.1  | FGFR2   | KGF receptor           |
|    | NM_006017.1  | CD133   | prominin-like 1       |    | NM_003012.2  | SFRP1   | frizzled-relat. prot. 1|
|    | NM_003506.1  | FZD6    | frizzled homolog 6    |    | NM_000165.2  | GJA1    | connexin 43            |
|    | NM_000165.2  | GJA1    | connexin 43           |    |              |         |                        |
|    | NM_005631.1  | SMO     | smoothened            |    |              |         |                        |
|    | NM_003107.1  | SOX4    | SRY-box 4             |    |              |         |                        |
|    | BF508662     | SPRY1   | sprouty 1             |    |              |         |                        |
|    | U54826.1     | MADH1   | SMAD1                 |    |              |         |                        |
|    | AK027071.1   | TSC22   | TGFβ-stimulated prot. |    |              |         |                        |
|    |              |         |                       |    |              |         |                        |
| H3 | AF268613.1   | POU5F1  | OCT4                  | K3 | AF268613.1   | POU5F1  | OCT4                   |
|    | NM_024674.1  | LIN-28  | RNA-binding protein   |    | NM_024674.1  | LIN-28  | RNA-binding protein    |
|    | NM_003212.1  | TDGF1   | Cripto                |    | NM_003212.1  | TDGF1   | Cripto                 |
|    | NM_024865.1  | NANOG   | ES transcription factor |  | NM_024865.1  | NANOG   | ES transcription factor|
|    | NM_003240.1  | EBAF    | left-right determ. fact. A |  | NM_003240.1 | EBAF  | left-right determ. fact. A |
|    | NM_020997.1  | LEFTB   | left-right determ. fact. B |  | NM_020997.1 | LEFTB | left-right determ. fact. B |
|    | NM_003577.1  | UTF1    | ES transcription factor 1 |  | NM_003577.1  | UTF1    | ES transcription factor 1 |
|    | AA218868     | THY1    | Thy-1 cell surface ag.|    | AA218868     | THY1    | Thy-1 cell surface ag. |
|    | NM_001290.1  | LDB2    | LIM domain binding 2  |    | NM_001290.1  | LDB2    | LIM domain binding 2   |
|    | AB028869.1   | BIRC5   | survivin              |    | NM_001168.1  | BIRC5   | survivin               |
|    | NM_003923.1  | FOXH1   | forkhead box H1       |    | NM_003923.1  | FOXH1   | forkhead box H1        |
|    | NM_002006.1  | FGF2    | FGF2                  |    | NM_002006.1  | FGF2    | FGF2                   |
|    | AF202063.1   | FGFR4   | FGFR4                 |    | NM_002011.2  | FGFR4   | FGFR4                  |
|    | L07335.1     | SOX2    | SRY-box 2             |    | L07335.1     | SOX2    | SRY-box 2              |
|    | M13077.1     | ALPP    | alkaline phosphatase  |    | M13077.1     | ALPP    | alkaline phosphatase   |
|    | NM_016941.1  | DLL3    | delta-like 3          |    | NM_016941.1  | DLL3    | delta-like 3           |
|    | NM_005585.1  | MADH6   | SMAD6                 |    | NM_005585.1  | MADH6   | SMAD6                  |
|    | NM_001134.1  | AFP     | alpha-fetoprotein     |    | NM_001134.1  | AFP     | alpha-fetoprotein      |
|    | NM_007295.1  | BRCA1   | breast cancer 1       |    | AF005068.1   | BRCA1   | breast cancer 1        |
|    | X95152       | BRCA2   | breast cancer 2       |    | X95152       | BRCA2   | breast cancer 2        |
|    | U96136.1     | CTNND2  | δ catenin 2           |    | AF035302.1   | CTNND2  | δ catenin 2            |
|    | NM_017412.1  | FZD3    | frizzled homolog 3    |    | NM_017412.1  | FZD3    | frizzled homolog 3     |
| | NM_020634.1 | GDF3 | growth diff. factor 3 | | NM_020634.1 | GDF3 | growth diff. factor 3 |
| | U91903.1 | FRZB | frizzled-related protein | | NM_001463.1 | FRZB | frizzled-related protein |
| | NM_012259.1 | HEY2 | hairy/enh. of split 2 YPRW | | AF098951.2 | ABCG2 | ATP-bind. cassette G2 |
| | U43148.1 | PTCH | patched | | | | |
| | U77914.1 | JAG1 | jagged 1 | | | | |
| | NM_004821.1 | HAND1 | heart and neural crest 1 | | | | |
| | NM_003392.1 | WNT5A | development regulator | | | | |
| | | | | | | | |
| H4 | NM_014676.1 | PUM1 | pumilio 1 | K4 | NM_005620.1 | S100A11 | calgizzarin |
| | D87078.2 | PUM2 | pumilio 2 | | NM_002965.2 | S100A9 | calgranulin B |
| | BC005912.1 | FCER1A | Fc frag. of IgE, high aff. I | | NM_002966.1 | S100A10 | calpactin I |
| | NM_019102.1 | HOXA5 | homeo box A5 | | NM_005978.2 | S100A2 | CAN19 |
| | BC005332.1 | IGKC | Ig const. κ | | NM_003125.1 | SPRR1B | cornifin |
| | BG340548 | IGHM | Ig heavy const. m | | NM_002203.2 | ITGA2 | integrin α2 |
| | NM_005574.2 | LMO2 | LIM domain only 2 | | NM_005547.1 | IVL | involucrin |
| | AA573862 | HLA-A | MHC I, A | | NM_005046.1 | KLK7 | kallikrein 7 |
| | X76775 | HLA-DMA | MHC II, DM α | | M19156.1 | KRT10 | keratin 10 |
| | | | | | X57348 | SFN | stratifin |
| | | | | | BC000125.1 | TGFB1 | TGFβ1 |
| | | | | | | | |
| H5 | NM_001738.1 | CA1 | carbonic anhydrase I | K5 | AL356504 | FLG | filaggrin |
| | NM_000129.2 | F13A1 | coag. factor XIII, A1 | | AF243527 | KLK5 | kallikrein 5 |
| | U62027.1 | C3AR1 | compl. comp. C3a R1 | | NM_006121.1 | KRT1 | keratin 1 |
| | AF130113.1 | CYB5-M | cytochrome b5 prec. | | NM_002274.1 | KRT13 | keratin 13 |
| | NM_001978.1 | EPB49 | eryth. memb. prot. 4.9 | | NM_000427.1 | LRN | loricrin |
| | NM_002000.1 | FCAR | Fc frag. of IgA | | NM_002963.2 | S100A7 | psoriasin 1 |
| | NM_004106.1 | FCER1G | Fc frag. of IgE, high aff. I | | NM_003238.1 | TGFB2 | TGFβ2 |
| | NM_004107.1 | FCGRT | Fc frag. of IgG, α | | NM_004245.1 | TGM5 | transglutaminase 5 |
| | NM_005143.1 | HP | haptoglobin | | | | |
| | NM_000558.2 | HBA1 | hemoglobin α1 | | | | |
| | H53689 | IGL@ | Ig l locus | | | | |
| | BE138825 | HLA-F | MHC I, F | | | | |
| | NM_002120.1 | HLA-DOB | MHC II, DO β | | | | |
| | | | | | | | |
| H6 | M81104.1 | CD34 | CD34 antigen | K6 | NM_001717.1 | BNC | basonuclin |
| | NM_005180.1 | BMI1 | B lymph. MLV ins. reg. | | AF091627.1 | TP73L | p63 |
| | NM_000222.1 | KIT | SCF receptor | | NM_002204.1 | ITGA3 | integrin α3 |
| | NM_005424.1 | TIE | endothelial RTK | | NM_000213.1 | ITGB4 | integrin β4 |

| | | | | | | |
|---|---|---|---|---|---|---|
| NM_003189.1 | TAL1 | SCL | | NM_001723.1 | BPAG1 | bullous pemph. ag.1 |
| D43968.1 | RUNX1 | RUNT TF 1 | | NM_000494.1 | BPAG2 | collagen XVII α1 |
| AL134303 | EGFL3 | EGF-like-domain 3 | | NM_000227.1 | LAMA3 | laminin α3 |
| NM_018951.1 | HOXA10 | homeo box A10 | | | | |

**Table 4. Select 20 Significant Genes in Human Embryonic Stem Cells[i].**

| Gene Symbol | Present in (6) | Title | ESC Signal | Fold change* |
|---|---|---|---|---|
| ZIC1 | | Zic finger transcription factor | 1011.53 | 257.85 |
| TDGF1 | X | teratocarcinoma-derived growth factor 1 | 1494.87 | 194.45 |
| CRABP1 | X | retinoic acid binding protein | 4379.50 | 138.00 |
| LIN-28 | X | RNA-binding protein LIN-28 | 1759.90 | 102.78 |
| NANOG | X | hypothetical protein FLJ12581 | 394.10 | 85.91 |
| FLJ10884 | | hypothetical protein FLJ10884 | 1707.63 | 64.11 |
| NLGN | | neuroligin | 222.60 | 51.72 |
| TMSNB | | thymosin, beta | 1550.60 | 46.26 |
| POU5F1 | X | OCT4 | 1641.17 | 35.51 |
| CYP26A1 | X | cytochrome P450, family member | 632.47 | 27.69 |
| FGF13 | | fibroblast growth factor 13 | 504.33 | 23.66 |
| SOX2 | X | SRY (sex determining region Y)-box 2 | 340.17 | 23.31 |
| LEFTB | X | left-right determination | 749.30 | 20.77 |
| FOXH1 | | forkhead box H1 | 248.93 | 19.15 |
| FZD7 | | frizzled homolog 7 | 996.23 | 16.65 |
| CDC6 | | cell division cycle 6 | 295.43 | 12.24 |
| REPRIMO | | mediator of the p53-dependent G2 arrest | 240.87 | 10.74 |
| PRODH | | proline dehydrogenase (oxidase) 1 | 353.63 | 9.24 |
| SPS | X | selenophosphate synthetase | 902.83 | 6.51 |
| SMARCA1 | | SWI/SNF related | 448.03 | 6.27 |